\documentclass[%
 amsmath,amssymb,
pra,
reprint,
dvipdfmx,
superscriptadress
]{revtex4-1}

\usepackage{graphicx}
\usepackage{dcolumn}
\usepackage{bm}
\usepackage{braket}

\usepackage{mathptmx}

\begin{document}

\preprint{APS/123-QED}


\title{Tracking Quantum Error Correction}
\author{Kosuke Fukui} \author{Akihisa Tomita} \author{Atsushi Okamoto} \affiliation{Graduate School of Information Science and Technology, \\Hokkaido University, Kita14-Nishi9, Kita-ku, Sapporo 060-0814, Japan}

\begin{abstract}
To implement fault-tolerant quantum computation with continuous variables, the Gottesman--Kitaev--Preskill (GKP) qubit has been recognized as an important technological element. We have proposed a method to reduce the required squeezing level to realize large scale quantum computation with the GKP qubit [Phys. Rev. X. {\bf 8}, 021054 (2018)], harnessing the virtue of analog information in the GKP qubits. In the present work, to reduce the number of qubits required for large scale quantum computation, we propose the tracking quantum error correction, where the logical-qubit level quantum error correction is partially substituted by the single-qubit level quantum error correction. In the proposed method, the analog quantum error correction is utilized to make the performances of the single-qubit level quantum error correction almost identical to those of the logical-qubit level quantum error correction in a practical noise level. The numerical results show that the proposed tracking quantum error correction reduces the number of qubits during a quantum error correction process by the reduction rate $\left\{{2(n-1)\times4^{l-1}-n+1}\right\}/({2n \times 4^{l-1}})$ for $n$-cycles of the quantum error correction process using the Knill's $C_{4}/C_{6}$ code with the concatenation level $l$.
Hence, the proposed tracking quantum error correction has great advantage in reducing the required number of physical qubits, and will open a new way to bring up advantage of the GKP qubits in practical quantum computation.
\end{abstract}
\maketitle
\section{Introduction}\label{Intro}
Quantum computation has a great deal of potential to efficiently solve some hard problems for conventional computers~\cite{Shor,Grov}. Among the candidates for qubits, squeezed vacuum states in an optical system have shown great potential for large scale continuous variable quantum computation; in fact, more than one million-mode continuous variable cluster state has already been achieved in an experiment~\cite{Yoshi}. This ability of entanglement generation comes from the fact that squeezed vacuum states can be entangled with only beam splitter coupling through the time-domain multiplexing approach to miniaturize optical circuits~\cite{Meni4,Meni5}. 
Moreover, a frequency-encoded continuous variable in an optical setup is also a promising quantum system to generate large scale cluster state more than thousands \cite{Phys,Chen,Ros}. 
In addition, promising platforms to implement large scale quantum computation are recently proposed in several setups such as a circuit QED \cite{Gri, Ter}, opto-mechanics \cite{Schm,Hou}, atomic ensembles \cite{Iked, Meni2}, and a trapped ion mechanical oscillator \cite{Fluh, Fluh2}.
However, since quantum computation with continuous variables itself has an analog nature, it is difficult to handle the accumulation of analog errors caused, for example, by photon loss during quantum computation~\cite{Ohl,Ohl2}. This can be circumvented by digitizing continuous variables using an appropriate code, such as the Gottesman--Kitaev--Preskill (GKP) code~\cite{GKP}, which are referred to as GKP qubits. 
Moreover, GKP qubits inherit the advantage of squeezed vacuum states on optical implementation; they can be entangled only by beam splitter coupling. 
Furthermore, we have proposed a high-threshold fault-tolerant quantum computation to alleviate the required squeezing level for fault-tolerant quantum computation to 9.8 dB~\cite{KF2}, and have taken a step closer to the realization of large scale quantum computation. Hence, the GKP qubits will play an indispensable role in implementing fault-tolerant quantum computation with continuous variables. 

In general, the quantum error correction (QEC) is repeatedly performed only by the logical-qubits during the quantum computation process. In large scale quantum computation, a large number of physical qubits are needed to obtain the highly accurate results of quantum computation. This required number of physical qubits is one issue that we should struggle with to implement large scale quantum computation. In this work, we propose a method to reduce the number of qubits required for the QEC during large scale quantum computation, where the logical-qubit level QEC is partially substituted by the single-qubit level QEC. 
The single-qubit level QEC~\cite{GKP} enables us to correct a displacement (deviation) error occurred in the single qubit by using a single ancilla qubit, unless the logical-qubit level error occurs.
Since the single-qubit level QEC can not correct the qubit-level errors, the bit and phase flip errors, we just $track$ the measurement outcomes in the single-qubit level QEC. Then, the QEC is performed with the help of a set of $tracked$ measurement outcomes in the single-qubit level QEC to correct the qubit-level errors. Although the single-qubit level QEC can be also implemented by discrete variables, the tracking QEC in discrete variables can not work well as shown later in the numerical results. By contrast, in our method, since the analog QEC makes the performances of the single-qubit level QEC almost identical to those of the logical-qubit level QEC, the tracking QEC can work well. The numerical results show that the proposed method has a great advantage to reduce the required number of qubits, e.g. in the concatenated QEC with analog QEC proposed in Ref. \cite{KF1}.

The rest of the paper is organized as follows. In Sec. \ref{Sec2}, we briefly review the background knowledge regarding the GKP qubit. In Sec. \ref{Sec3}, we propose the method to reduce the number of qubits required for large scale quantum computation.  In Sec. \ref{Sec4}, the numerical results show the superiority of the proposed method over the conventional methods. Section \ref{Sec5} is devoted to discussion and conclusion.
\section{Preliminaries}\label{Sec2}
In this section, we review the some background knowledge regarding the GKP qubit, a noise model considered in this work, the single-qubit level QEC, and the analog QEC.
\subsection{The GKP qubit}\label{Sec2A}
Gottesman, Kitaev, and Preskill proposed a method to encode a qubit in an oscillator's $q$ (position) and $p$ (momentum) quadratures to correct errors caused by a small deviation in the $q$ and $p$ quadratures~\cite{GKP}. This error correction of a small deviation can handle any error acting on the oscillator, even a superposition of displacements. 

The basis of the GKP qubit is composed of a series of Gaussian peaks of width $\sigma$ and separation $\sqrt{\pi}$ embedded in a larger Gaussian envelope of width 1/$\sigma$. Although in the case of infinite squeezing ($\sigma \rightarrow 0$) the GKP qubit bases become orthogonal, in the case of finite squeezing, the approximate code states are not orthogonal. The approximate code states $\ket {\widetilde{0}}$ and $\ket {\widetilde{1}}$ are defined as  
\begin{eqnarray}
\ket {\widetilde{0}} &\propto &   \sum_{t=- \infty}^{\infty} \int \mathrm{e}^{-2\pi\sigma^2t^2}\mathrm{e}^{-(q-2t\sqrt{\pi})^2/(2\sigma^2)}\ket{q}  dq,  \\
\ket {\widetilde{1}} &\propto & \sum_{t=- \infty}^{\infty} \int \mathrm{e}^{-\pi\sigma^2(2t+1)^2/2}  
\mathrm{e}^{-(q-(2t+1)\sqrt{\pi})^2/(2\sigma^2)}\ket{q}  dq .     
\end{eqnarray}
 In the case of the finite squeezing, there is a finite probability of misidentifying $\ket {\widetilde{0}}$ as $\ket {\widetilde{1}}$, and vice versa. Provided the magnitude of the true deviation is less than $\sqrt{\pi}/2$ from the peak value, the decision of the bit value from the measurement of the GKP qubit is correct. The probability $p_{\rm corr}$ to identify the correct bit value is the area of a normalized Gaussian of a variance ${{\sigma}}^2$ that lies between $-\sqrt{\pi}/2$ and $\sqrt{\pi}/2$~\cite{Meni}:
\begin{equation}
p_{\rm corr} = \int_{\frac{-\sqrt{\pi}}{2}}^{\frac{\sqrt{\pi}}{2}} dx \frac{1}{\sqrt{2\pi {\sigma} ^2}} {\rm exp}(-x^2/{2{\sigma} ^2}).
\label{eq3}
\end{equation}

In addition to the imperfection that originates from the finite squeezing of the initial states, we consider the degradation in the Gaussian quantum channel~\cite{GKP,Harri}, which leads to a displacement in the quadrature during the quantum computation. The channel is described by a superoperator $\zeta $ acting on a density operator $\rho$ as follows: 
\begin{equation}
\rho \to \zeta (\rho) = \frac{1}{\pi{\xi }^2}\int d^2\alpha \mathrm{e}^{-{| \alpha |}^2/{{\xi}^2}}D( \alpha ) \rho D( \alpha ) ^{\dagger },
\label{eq4}
\end{equation}
where $D(\alpha)$ is a displacement operator in the phase space. The position $q$ and momentum $p$ are displaced by $ D( \alpha )$ independently as
\begin{equation}
q \to q + v, \hspace{10pt}
p \to p + u, 
\end{equation}
respectively, where $v$ and $u$ are real Gaussian random variables with mean zero and variance $\xi ^2$. From Eq. \ref{eq4}, we see that the Gaussian quantum channel conserves the position of the Gaussian peaks in the probability density function on the measurement outcome of the GKP qubit, but increases the variance as 
\begin{equation}
\sigma^2  \to \hspace{5pt} \sigma^2 \ +\hspace{5pt} \xi^2, 
\end{equation}
where the $\sigma^2$ is the variance before the Gaussian quantum channel. Therefore, in the next section, we evaluate the performance of a QEC method with a code capacity noise model, where the noise is parameterized by a single variance $\sigma^2$ that includes the squeezing level of the initial GKP qubit and the degradation via the Gaussian quantum channel.
\subsection{The single-qubit level quantum error correction}\label{Sec2B}
In Ref.~\cite{GKP}, the single-qubit level QEC has been proposed to correct a displacement (deviation) error derived from the finite squeezing of the GKP qubit or the Gaussian quantum channel. We here explain the single-qubit level QEC to correct the displacement error in the $p$ quadrature in detail. In this single-qubit level QEC in the $p$ quadrature, an additional single ancilla qubit is entangled with the data qubit by a CNOT gate, where the data qubit is the target qubit. The ancilla qubit is prepared in the state $\ket {\widetilde{0}}$ to prevent us from identifying the bit value of the data qubit. The CNOT gate, which corresponds to the operator exp(-$i\hat{q}_{\rm a}\hat{p}_{\rm D}$) for continuous variables, transforms
\begin{eqnarray}
 \hat{q}_{\rm a} &\to &   \hat{q}_{\rm a}, \\
 \hat{p}_{\rm a} &\to & \hat{p}_{\rm a} - \hat{p}_{\rm D} ,  \\
 \hat{q}_{\rm D} &\to &  \hat{q}_{\rm D}+ \hat{q}_{\rm a}, \\
 \hat{p}_{\rm D} &\to &  \hat{p}_{\rm D},
\end{eqnarray}
where $\hat{q}_{\rm D} (\hat{p}_{\rm D})$ and $\hat{q}_{\rm a} (\hat{p}_{\rm a})$ are the quadrature operators of the data and ancilla qubits in the position $q$ (momentum $p$), respectively. 
Regarding the deviation, the CNOT gate operation displaces the deviation of the $q$ and $p$ quadratures as
\begin{eqnarray}
\overline{\Delta}_{\rm {\it q},a} &\to & \overline{\Delta}_{\rm {\it q},a} ,  \\
 \overline{\Delta}_{\rm {\it p},a} &\to &\overline{\Delta}_{\rm {\it p},a}- \overline{\Delta}_{\rm {\it p},D}, \\
\overline{\Delta}_{\rm {\it q},D} &\to  &\overline{\Delta}_{\rm {\it q},D}+ \overline{\Delta}_{\rm {\it q},a} , \\
\overline{\Delta}_{\rm {\it p},D} &\to &\overline{\Delta}_{\rm {\it p},D},
\end{eqnarray}
where $\overline{\Delta}_{\rm {\it q},D} ( \overline{\Delta}_{\rm {\it p},D})$ and $\overline{\Delta}_{\rm {\it q},a}  (\overline{\Delta}_{\rm {\it p},a} )$ are the true deviation values of the data and ancilla qubits in the position $q$ (momentum $p$), respectively.
{We assume that the deviations of the data qubit in the $q$ and $p$ quadratures obey the Gaussian distribution with the variance $\sigma^2_{{\rm D},q}$ and $\sigma^2_{{\rm D},p}$, and the deviations of the {ancilla qubit} in the $q$ and $p$ quadratures {obey} the Gaussian distribution with the variance{$\sigma^2_{{\rm a},q}$ and $\sigma^2_{{\rm a},p}$, respectively.}
After the CNOT gate, we measure the ancilla qubit in the $p$ quadrature, and obtain the deviation of the ancilla qubit ${\Delta}_{\rm m{\it p}, a}$ 
{that obeys the Gaussian distribution with the variance $\sigma^2_{{\rm D},p}+{\sigma^2_{{\rm a},p}}$.
Then, we perform the displacement $|{\Delta}_{\rm m{\it p}, a}|$ on the $p$ quadrature of the data qubit to correct by shifting back in the direction to minimize the deviation. }
If $|{\Delta}_{\rm  m{\it p}, a}| = |\overline{\Delta}_{\rm {\it p},a}- \overline{\Delta}_{\rm {\it p},D}|$ is less than $\sqrt{\pi}/2$, the true deviation value of the data qubit in the $p$ quadrature changes from $\overline{\Delta}_{\rm {\it p},D}$ to $ \overline{\Delta}_{\rm {\it p},a}$ after the displacement operation, which displaces $\overline{\Delta}_{\rm {\it p},D}$ by ${\Delta}_{\rm  m{\it p}, a}(=\overline{\Delta}_{\rm {\it p},a}- \overline{\Delta}_{\rm {\it p},D})$. On the other hand, if $ |\overline{\Delta}_{\rm {\it p},a}- \overline{\Delta}_{\rm {\it p},n}|$ is more than $\sqrt{\pi}/2$, the bit error in the $p$ quadrature occurs after the displacement operation. 
Therefore, the single-qubit level QEC for the data qubit in the $p$ quadrature can reduce the variance of the data qubit in the $p$ quadrature from $\sigma^2_{{\rm D},p}$ to ${\sigma^2_{{\rm a},p}}$. The variance of the data qubit in the $q$ quadrature after the single-qubit level QEC increases from $\sigma^2_{{\rm D},q}$ to $\sigma^2_{{\rm D},q}+{\sigma^2_{{\rm a},q}}$, since the true deviation $\overline{\Delta}_{\rm {\it q},D}$ and $ \overline{\Delta}_{\rm {\it q},a}$ obey the Gaussian distribution with the variance $\sigma^2_{{\rm D},q}$ and ${\sigma^2_{{\rm a},q}}$, respectively, where the $\overline{\Delta}_{\rm {\it q},D}$ and $\overline{\Delta}_{\rm {\it q},a}$ are the true deviation of the data qubit and the ancilla qubit, respectively. 

Similarly, the single-qubit level QEC in the $q$ quadrature can be performed using the second ancilla qubit, where the ancilla is prepared in the state $\ket {\widetilde{+}}$ and the data qubit is assumed to be the control qubit.
Regarding the deviation, the CNOT gate operation displaces the deviation of the $q$ and $p$ quadratures as
\begin{eqnarray}
\overline{\Delta}_{\rm {\it q},a2} &\to & \overline{\Delta}_{\rm {\it q},a2}+ \overline{\Delta}_{\rm {\it q},D} + \overline{\Delta}_{\rm {\it q},a} ,  \\
 \overline{\Delta}_{\rm {\it p},a2} &\to &\overline{\Delta}_{\rm {\it p},a2}, \\
\overline{\Delta}_{\rm {\it q},D} + \overline{\Delta}_{\rm {\it q},a}&\to  &\overline{\Delta}_{\rm {\it q},D} + \overline{\Delta}_{\rm {\it q},a}, \\
\overline{\Delta}_{\rm {\it p},a} &\to &\overline{\Delta}_{\rm {\it p},a}-\overline{\Delta}_{\rm {\it p},a2},
\end{eqnarray}
where $\overline{\Delta}_{\rm {\it q},a2} ( \overline{\Delta}_{\rm {\it p},a2})$ is the true deviation value of the second ancilla qubit in the position $q$ (momentum $p$).
After the CNOT gate, we measure the ancilla qubit in the $q$ quadrature, and obtain the deviation of the ancilla qubit ${\Delta}_{\rm m{\it q}, a2}$ . 
Then, we perform the displacement $|{\Delta}_{\rm m{\it q}, a2}|$ on the $q$ quadrature of the data qubit to correct by shifting back in the direction to minimize the deviation. 
If $|{\Delta}_{\rm  m{\it q}, a2}| = |\overline{\Delta}_{\rm {\it q},a2}+ \overline{\Delta}_{\rm {\it q},D} + \overline{\Delta}_{\rm {\it q},a}|$ is less than $\sqrt{\pi}/2$, the true deviation value of the data qubit in the $q$ quadrature changes from $\overline{\Delta}_{\rm {\it q},D} + \overline{\Delta}_{\rm {\it q},a}$ to $ -\overline{\Delta}_{\rm {\it q},a2}$ after the displacement operation, which displaces $\overline{\Delta}_{\rm {\it q},D} + \overline{\Delta}_{\rm {\it q},a}$ by $-{\Delta}_{\rm  m{\it p}, a}(=-\overline{\Delta}_{\rm {\it q},a2}- \overline{\Delta}_{\rm {\it q},D} - \overline{\Delta}_{\rm {\it q},a})$. On the other hand, if $|{\Delta}_{\rm  m{\it q}, a2}|$ is more than $\sqrt{\pi}/2$, the bit error in the $q$ quadrature occurs after the displacement operation. 
To summarize, after the sequential single-qubit level QECs in the $p$ and $q$ quadrature, the variances of the data qubit in the $q$ and $p$ quadratures become ${{\sigma}_{{\rm a},q}}^2$ and $2{{\sigma}_{{\rm a},p}}^2$}, respectively. Although the single-qubit level QEC works well for the small deviation, we need to operate the logical-qubit level QEC to correct the deviation greater than $\sqrt{\pi}/2$. 

\subsection{Analog quantum error correction}\label{Sec2C}
\begin{figure}[t]
 \includegraphics[angle=0, width=1.0\columnwidth]{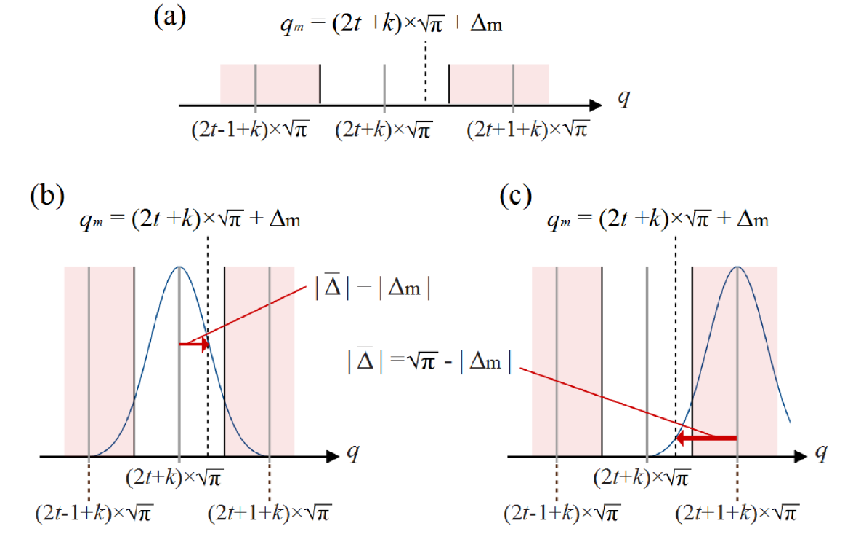}
     \setlength\abovecaptionskip{7pt}
 \caption{\label{Fig1}Introduction of a likelihood function. (a) Measurement outcome and deviation from the peak value in $q$ quadrature. The dotted line shows the measurement outcome $q_{\rm m}$ equal to $(2 t + k)\sqrt{\pi}+\Delta_{\rm m}$ $(t = 0, \pm 1, \pm 2,\cdots,\ k = 0, 1)$, where $k$ is defined as the bit value that minimizes the deviation $\Delta_{\rm m}$. The red areas indicate the area that yields code word $(k+1)$ mod 2, whereas the white area denotes the area that yields the codeword $k$. (b) and (c) Gaussian distribution functions as likelihood functions of the true deviation $\overline{\Delta}$ represented by the arrows. (b) refers to the case of the correct decision, where the amplitude of the true deviation is $|\overline{\Delta}| < \sqrt{\pi}/2$, whereas (c)  the case of the incorrect decision $ \sqrt{\pi}/2 < |\overline{\Delta}| < \sqrt{\pi}$.}
 \label{fig1}
 \end{figure}

We explain the analog QEC, which improves the performance of the tracking QEC (see also~\cite{KF1} for details of the analog QEC). 
In the measurement of the GKP qubit for the computational basis, we make a decision on the bit value $k ( = 0, 1)$ from the measurement outcome of the GKP qubit in the $q$ quadrature $q_{\rm m}= q_{k}+ \Delta_{\rm m}$ to minimize the deviation $|\Delta_{\rm m}|$, where $q_{k}$ $(k = 0,1)$ is defined as $(2 t + k)\sqrt{\pi}$ $(t = 0, \pm 1, \pm 2,\cdots.)$ as shown in Fig.~\ref{fig1} (a). 
{The decision is correct when the magnitude of the true deviation $|\overline{\Delta}|$ satisfies $|\overline{\Delta}|<\sqrt{\pi}/2$ as shown in Fig. \ref{fig1} (b), while it is incorrect when $\sqrt{\pi}/2<|\overline{\Delta}|<3\sqrt{\pi}/2$.}
In the digital QEC~\cite{Pou,Goto}, we obtain the bit value from analog outcome and calculate the likelihood from only the binary information regardless of the value of $\Delta_{\rm m}$, since we consider the GKP qubit as a qubit. The likelihood of the correct decision is calculated by $p_{\rm corr}$ in Eq.~(\ref{eq3}) using bit value and the noise level ${\sigma} ^2$.
The likelihood of the incorrect decision is calculated by $1- p_{\rm corr}$. On the contrary, we consider $\Delta_{\rm m}$ in the analog QEC. We employ a Gaussian function, which the true deviation $|\overline{\Delta}|$ obeys, as a likelihood function. The likelihood of the correct decision is calculated by
\begin{equation}
 f(\overline{\Delta}) = f(\Delta_{\rm m}) = \frac{1}{\sqrt{2\pi\sigma^{2}}} \mathrm{e}^{-\overline{\Delta}^{2}/(2\sigma^{2})}.
 \label{eq15}
 \end{equation}
 The likelihood of the incorrect decision is calculated by 
 \begin{equation}
 f(\overline{\Delta}) =f(\sqrt{\pi}-|\Delta_{\rm m}|).
\label{eq16}
\end{equation}
Strictly speaking, the likelihood function should be a periodic function including the sum of the Gaussian functions, considering that the GKP state is the superposition of the Gaussian states. Nevertheless, in this paper, the likelihood function is approximated by simple Gaussian functions given by Eqs. (\ref{eq15}) and (\ref{eq16}), since the tail of the Gaussian function second nearest to the measurement outcome is small enough to ignore.
In the QEC, we can reduce the decision error on the entire code word by considering the likelihood of the joint event of multiple qubits to choose the most likely candidate. 
As a result, the analog QEC under the code capacity model can improve the QEC performance with a single block code without the concatenation such as the three-qubit flip code~\cite{KF1}. In the previously proposed digital QEC~\cite{Pou,Goto} has been shown to improve the QEC performance with only the concatenated code.
 \section{The tracking quantum error correction}\label{Sec3}
 \subsection{Logical-qubit level quantum error correction}\label{Sec3A}
To implement large scale quantum computation, a number of single (physical) qubits should be encoded into a logical qubit to correct errors on the logical qubit. Then, by using a fault-tolerant manner such as a concatenation, the failure probability of the logical-qubit level QEC can be reduced to an arbitrary value, if the error probability on a physical qubit is less than the threshold value, which varies on the QEC code. Since the logical-qubit level QEC is repeated during the quantum computation process, a large number of physical qubits are needed to obtain highly accurate results on the quantum computation. For example, for the Knill's $C_{4}/C_{6}$ code~\cite{Knill} with the concatenation level $k$, the required number of physical qubits to prepare a logical qubit and a Bell pair with level $l$ are $4\times12^{l-1}$ and $16\times12^{l-1}$, respectively, where the logical qubit is composed of $4\times3^{l-1}$ physical qubits. Accordingly, this required number of physical qubits for the logical-qubit level QEC is one issue that we should struggle with to implement large scale quantum computation. 

 \subsection{Tracking quantum error correction}\label{Sec3B}
\begin{figure}[t]
 \includegraphics[angle=0, width=1.0\columnwidth]{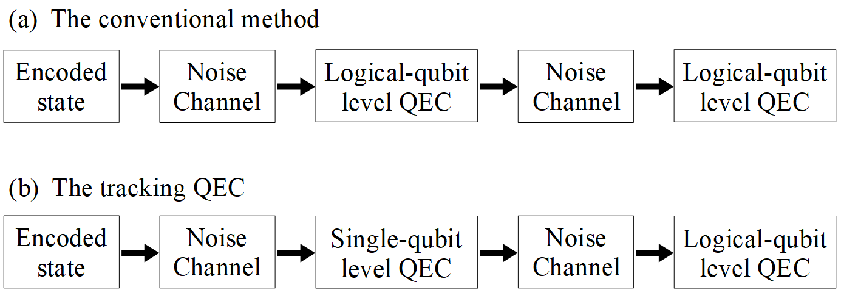}
     \setlength\abovecaptionskip{15pt}
 \caption{Introduction of the tracking QEC. (a) The conventional QEC with the two-QEC cycle, where the QECs are performed with the only logical-qubit level QEC. (b) The tracking QEC with the two-QEC cycle, where the first logical-qubit level QEC in the conventional method is substituted by the single-qubit level QEC. }
 \label{fig2}
\end{figure}

In general, the QEC is repeatedly performed only in the logical-qubit level during the quantum computation process as shown in Fig. \ref{fig2} (a). We propose a method to reduce the required number of qubits, which we call "tracking QEC", because the logical-qubit level QEC is partially substituted by the single-qubit level QEC~\cite{GKP} whose measurement outcome is tracked in the repeated QEC process as shown in Fig. \ref{fig2} (b). In our method, we apply analog QEC~\cite{KF1} to the tracking QEC to improve the performance. Since the single-qubit level QEC can reduce the error probability as described in Sec.~\ref{Sec2B} and the number of qubits required for the single-qubit level QEC is less than that for the logical-qubit level one, the substitution will reduce the required number of qubits.

To provide an insight into our method, we focus on the tracking QEC with the two-QEC cycle, where the QEC after the Gaussian quantum channel is repeated twice as shown in Fig. \ref{fig2}. As a specific example of a QEC code, we use the Knill's $C_{4}/C_{6}$ code~\cite{Knill}, where the error correction is based on quantum teleportation (see also Appendixes A and B for details of with the conventional and proposed methods using the $C_{4}/C_{6}$ code). The quantum teleportation process refers the outcomes $M_{p}$ and $M_{q}$ of the Bell measurement on the encoded qubits, and determines the transformation of the qubits. We obtain the Bell measurement outcomes of bit values $m_{pi}$ and $m_{qi}$ for the $i$-th physical GKP qubit of the encoded data qubit and the encoded Bell state, respectively. 
In addition to bit values, we also obtain deviation values $\Delta_{p{\rm m}i}$ and $\Delta_{q{\rm m}i}$ for the $i$-th physical GKP qubit. In our method, the first and second QECs are performed by the single- and logical-qubit level QECs, respectively. Since the single-qubit level QEC can not correct the qubit-level error, we just $track$ the measurement outcomes in the first QEC. 
After the two QECs, we obtain a set of the likelihoods from the results of the first and second QECs. From the set of the likelihoods, we consider the likelihood of the following the two possible joint events: one is the correct decision, where no qubit-level error occurs in both QECs. In this case, both true deviation values of the first and second QECs, $|\overline{\Delta }^{(1)}|$ and $|\overline{\Delta }^{(2)}|$, are less than $\sqrt{\pi}/2$ or more than $\sqrt{\pi}/2$. 
When both true deviation values are less than $\sqrt{\pi}/2$, $|\overline{\Delta }^{(1)}|$ and $|\overline{\Delta }^{(2)}|$ are equal to $|\Delta_{\rm m}^{(1)}|$ and $|\Delta_{\rm m}^{(2)}|$, respectively. 
When both true deviation values are more than $\sqrt{\pi}/2$, $|\overline{\Delta }^{(1)}|$ and $|\overline{\Delta }^{(2)}|$ are equal to $\sqrt{\pi}-|\Delta_{\rm m}^{(1)}|$ and $\sqrt{\pi}-|\Delta_{\rm m}^{(2)}|$, respectively. 
The other is the incorrect decision, where the single error occurs in either of the two QECs. 
In this case, one of the two true deviation values of the first and second QECs is greater than $\sqrt{\pi}/2$, and satisfies $|\bar{\Delta}^{(1)}|=|\Delta_{\rm m}^{(1)}|$ and $|\bar{\Delta}^{(2)}|+|\Delta_{\rm m}^{(2)}|=\sqrt{\pi}$, or $|\bar{\Delta}^{(1)}|+|\Delta_{\rm m}^{(1)}|=\sqrt{\pi}$ and $|\bar{\Delta}^{(2)}|=|\Delta_{\rm m}^{(2)}|$. 
Hence, the likelihoods for the correct decision without and with analog QEC are calculated by 
\begin{eqnarray}
F_{\rm corr} &=&{p_{\rm corr}}^{2}+(1-{p_{\rm corr}})^{2},\label{eq16}\\
F_{\rm corr}^{\rm ana}&=&f(|\Delta_{\rm m}^{(1)}|) f(|\Delta_{\rm m}^{(2)}|) \nonumber \\
&~&\hspace{10pt}+f(\sqrt{\pi}-|\Delta_{\rm m}^{(1)}|)f(\sqrt{\pi}-|\Delta_{\rm m}^{(2)}|),\label{eq17}
\end{eqnarray}
respectively, where $p_{\rm corr}$ is given by Eq. (\ref{eq3}). The likelihoods for the incorrect decision without and with analog QEC are calculated by
\begin{eqnarray}
F_{\rm in} &=& 2(1-p_{\rm corr}){p_{\rm corr}}, \hspace{50pt}\label{eq18}\\
F_{\rm in}^{\rm ana}&=&f(|\Delta_{\rm m}^{(1)}|)f(\sqrt{\pi}-|\Delta_{\rm m}^{(2)}|) \hspace{50pt} \nonumber \\
&~&\hspace{20pt}+f(\sqrt{\pi}-|\Delta_{\rm m}^{(1)}|)f(|\Delta_{\rm m}^{(2)}|), \label{eq19}
\end{eqnarray}
respectively. By considering these likelihoods of the joint event and choosing the most likely candidate, we can reduce the decision error on the entire code word after the second logical-qubit level QEC. By contrast, the single-qubit level QEC does not work, and QEC is performed by two independently operating logical-level QECs. Although we focus on the tracking QEC with the GKP qubits, we note that the tracking QEC with the discrete variables can be also performed, where the likelihoods are given by Eqs. (\ref{eq16}) and (\ref{eq18}). 
In our method, we utilize analog QEC using Eqs. (\ref{eq17}) and (\ref{eq19}) to make the performances of the single-qubit level QEC almost identical to that of the logical-qubit level QEC as shown in the numerical calculations.

We here estimate the required number of physical qubits to implement the two QECs. 
In the $C_{4}/C_{6}$ code with the concatenation level $l$ ($l\geq 1$), the logical qubit is composed of the $4\times 3^{l-1}$ physical qubits, and the preparation of the logical qubit and the logical Bell pair level $l$ consumed $4\times 12^{l-1}$ and $16\times12^{l-1}$, respectively \cite{Goto,Knill}. 
Therefore, the required number of physical qubits for the logical-qubit level QEC is $16\times12^{l-1}$, where the logical-qubit level QEC consumes the logical Bell pair. 
By contrast, the required number of physical qubits for the single-qubit level QEC is $2\times4\times 3^{l-1}$, where each physical qubit composing the logical data qubit consumes two ancilla physical qubits to correct the small deviation in the $q$ and $p$ quadratures. 
In the case of the QEC process with the two-QEC cycle, the number of the physical qubits for the conventional method $R_{\rm con}^{(2,l)}$ and proposed method  $R_{\rm pro}^{(2,l)}$ are 
\begin{eqnarray}
 R_{\rm con}^{(2,l)}&=&2 \times 16\times12^{l-1},\\
R_{\rm pro}^{(2,l)}&=&2 \times 4\times 3^{l-1}+16\times12^{l-1},
\end{eqnarray}
respectively. 
Hence, the proposed method for the two-QEC cycle reduces by $ R_{\rm con}^{(2,l)}- R_{\rm pro}^{(2,l)}$ = $16\times12^{l-1}-8\times3^{l-1}$ physical qubits with the concatenation level $l$. 
Similarly, the conventional and proposed methods for the $n$-QEC cycle, consume the physical qubits $R_{\rm con}^{(n,l)}$ and $R_{\rm con}^{(n,l)}$ as 
\begin{eqnarray}
R_{\rm con}^{(n,l)}&=&n \times 16\times12^{l-1},\\
R_{\rm pro}^{(n,l)}&=&2 (n-1) 4\times 3^{l-1}+16\times12^{l-1}, 
\end{eqnarray}
respectively.
Hence, the proposed method for the $n$-QEC cycle can reduce $ R_{\rm con}^{(n,l)}-R_{\rm pro}^{(n,l)}=(n-1)\times\left\{ R_{\rm con}^{(2,l)}-R_{\rm pro}^{(2,l)}\right\}$ physical qubits, where the single- and logical-qubit level QECs are performed in the first $(n-1)$-QECs and the $n$-th QEC, respectively. 
Finally, we describe the reduction rate of the number of physical qubits per $n$-QEC cycle. For the $n$-QEC cycle using $C_{4}/C_{6}$ code with the level $l$, the reduction rate is obtained by
\begin{equation}
\frac{R_{\rm con}^{(n,l)}-R_{\rm pro}^{(n,l)}}{R_{\rm con}^{(n,l)}} = \frac{2(n-1)\times 4^{l-1}-n+1}{2n\times 4^{l-1}}.
\end{equation}

\section{Numerical calculations}\label{Sec4}
To validate the effectiveness of our proposed method, we calculate the failure probability of the QEC and the number of physical qubits required in the QEC using the Monte Carlo method. We examine the tracking QEC performance, taking the conventional logical-qubit level QEC without the analog QEC as a reference. We simulate the QEC after the Gaussian quantum channel is repeated twice as described in Sec.~\ref{Sec3}. 
In this simulation, we use the Knill's $C_{4}/C_{6}$ code~\cite{Knill} for the concatenation and assume that the encoded data qubit, encoded Bell state, and the physical qubits are prepared perfectly, and the variance of the GKP qubits of the encoded data qubit is increased to $\sigma^2$ only by the Gaussian quantum channel. 
In the noise channel of the Gaussian quantum channel with $n$-cycles, the logical-qubit suffers from each of $n$-noise channels independently by the same amount of the noise.

\begin{figure}[t]
 \includegraphics[angle=0, width=1.0\columnwidth]{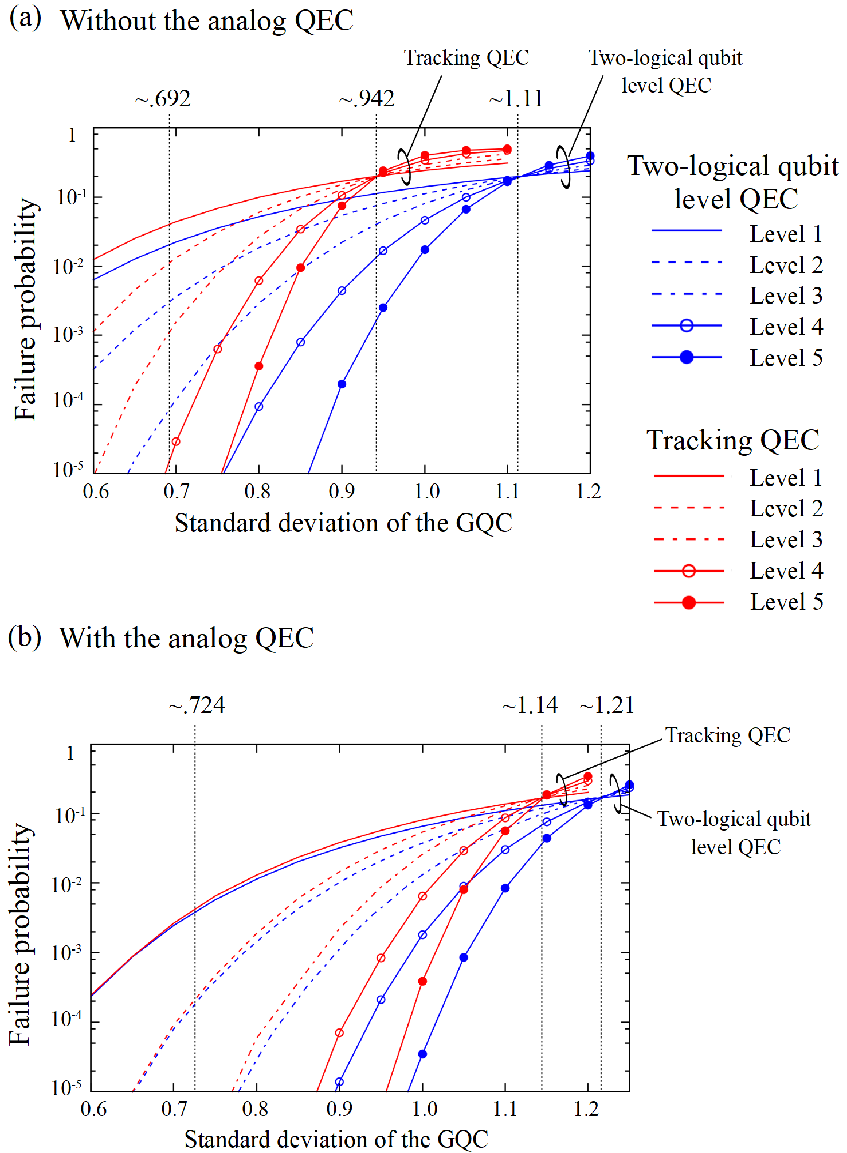}  
  \caption{Simulation results for the failure probabilities of the two QECs for the $q$ ($p$) quadrature with the $C_{4}/C_{6}$ code using the conventional (blue line) and proposed method (red line). 
GQC describes the Gaussian quantum channel. The results without the analog QEC (a) and with the analog QEC (b) are represented for the concatenation level 1 (solid), level 2 (dashed), level 3 (dashed-dotted), level 4 (open circles), and level 5 (filled circles), respectively. The thresholds are indicated by the vertical dashed lines with $\sim 1.11$, $\sim 0.942$, $\sim 1.21$, and $\sim 1.14$. The vertical dashed lines with $\sim 0.692$ and $\sim 0.724$ indicate the standard deviation for the practical noise level defined in the main text.
}
  \label{fig3}
\end{figure}

In Fig. \ref{fig3}, the failure probabilities for the $q$ ($p$) quadrature up to level 5 of the concatenation are plotted as a function of the noise level as the standard deviation of the Gaussian quantum channel. The noise is given by the sum of the noise of the first and second QECs, where the encoded state suffers from the same amount of the noise in the first and second QECs. 
We here define the threshold for the two-QEC cycle as;
if the sum of the noise of two cycles is below the threshold, the failure probability with the concatenation level $l$ can be reduced super exponentially as $l$ becomes large. Hence, all the lines plotted as a function of a noise level for various concatenation level $l$ meet at a single point, indicating the threshold value. 
In Fig. \ref{fig3}, we have plotted the failure probabilities (a) without the analog QEC and (b) with the analog QEC for the conventional two logical-qubit level QECs and the proposed method, respectively. 
 In the case of the Gaussian quantum channel, the displacement errors in the $q$ and $p$ quadrature occur independently. Considering the CSS code, where errors in the $q$ and $p$ quadrature can be treated separately, the failure probabilities for the $q$ and $p$ quadrature has the same value.
As shown in Fig. \ref{fig3}, the conventional method without and with the analog QECs achieve the threshold values of the standard deviation values $\sim 1.11$ and $\sim 1.21 $, respectively. 
These threshold values are identical to twice of those obtained for the single cycle of the logical-qubit level QEC $\sim 0.555$ and $\sim 0.607 $ in Ref.~\cite{KF1}, respectively. 
It results from the fact that the failure probability of the two-QEC cycle for the conventional method is calculated by $2\times P(1-P)$, where the probability $P$ is the failure probability of the single cycle of the logical-qubit level QEC.
This is because the sequential logical-qubit level QECs are performed independently after the each of two noise channels, and the event to fail in the logical-qubit level QEC occurs independently.

Fig.~\ref{fig3} (a) show that the tracking QEC degrades the threshold of the standard deviation by $\sim$ 0.17 without the analog QEC. Fig.~\ref{fig3} (b) show that the tracking QEC also degrades the threshold of the standard deviation by $\sim$ 0.07 with analog QEC. However, the degradation with the analog QEC is smaller than that without the analog QEC. 
More specifically, we compare the ratio of failure probabilities between the tracking QEC without and with the analog QEC at the same noise level of the standard deviation and the same concatenation level. For example, the ratio for the analog QEC with the concatenation level 1 is obtained by 0.00421/0.00375 $\sim 1.1$, similarly the ratios with the concatenation level 2 and 3 are $\sim 1.2$ and $\sim 1.4$, respectively, by contrast, the ratios without the analog QEC with the concatenation level 1, 2 and 3 are $\sim 1.8$, $\sim 3.7$ and $\sim 13.8$, respectively.
Hence, it is clear that the ratios of the proposed method with the analog QEC are greater than that without the analog QEC. 
In addition, it is remarkable that the tracking QEC with the analog QEC in Fig.~\ref{fig3} (b) suppresses errors more effectively than the conventional method without the analog QEC. Furthermore, it is also remarkable that the ratios of the tracking QEC with the analog QEC become greater as the noise level of the standard deviation become smaller, and the analog QEC makes the performances of the single-qubit level QEC almost identical to that of the logical-qubit level QEC in a low noise level. 
These results show the virtue of use of analog information. On the basis of these results, for the $2$-QEC cycle, we can conclude that the proposed method with analog QEC in the practical noise level can achieve efficient resource reduction by $16\times12^{l-1}-8\times3^{l-1}$ physical qubits with the concatenation level $l$ with only a small impact on the QEC performance, where the reduction rate for the $2$-QEC cycle is $({2\times4^{l-1}-1})/({4 \times 4^{l-1}})=1/2-1/(4\times4^{l-1})$. Hence, the reduction rate becomes close to 50 $\%$ for larger $l$, where the reduction rates are 25, 43.8, 48.4, 49.6 and 49.9 for the level 1, 2, 3, 4, and 5, respectively.

In the following, we consider admissible noise level of the Gaussian quantum channel for the tracking QEC. In practice, fault-tolerant quantum computation should be performed with a noise level smaller than the threshold value so as not to spend huge amounts of single qubits to prepare logical qubits with the required concatenation level $l$. To evaluate our proposed method, we assume that the single- and logical-qubit level QECs are performed with one-tenth of the threshold value according to Refs.~\cite{Dev, Jon}. 
For simplicity, we use the threshold value as the rate of the misidentifying the bit value of the GKP qubit. 
In the logical-qubit level QEC, the threshold of the noise level per cycle $\sim 0.555$ and $\sim 0.607 $ for without and with analog QEC correspond to the error rate of the misidentifying the bit value $\sim11.0$ $\%$ and  $\sim14.3$ $\%$, respectively, where the error rate of the misidentifying is obtained by $1-p_{\rm corr}$ given in Eq. (\ref{eq3}). Therefore, we set the rate of the misidentifying the bit value as $\sim1.1$ $\%$ and $\sim1.43$ $\%$ which correspond to a noise level $\sim 0.346$ and  $\sim 0.362$. As shown in Fig.~\ref{fig3} (a), there is a gap of failure probabilities between the conventional and proposed method with the set noise level of $\sim 2\times 0.346 = 0.692$ without the analog QEC. 
By contrast, the failure probabilities of the proposed method with the analog QEC is almost same as that of the conventional method with the set noise level $\sim 2\times 0.362 = 0.724$ as shown in Fig.~\ref{fig3} (b).

\section{Discussion and conclusion}\label{Sec5}
In this work, we have proposed the tracking QEC with analog QEC to reduce the number of qubits required for large scale quantum computation, bringing up the advantages of the GKP qubits in practical quantum computation. In the proposed method, the single-qubit level QEC is combined with the standard logical-qubit level QEC, in a way that a part of the logical-qubit level QEC is substituted by the single-qubit level QEC during the quantum computation. 
Furthermore, we propose to apply the analog QEC to the tracking QEC to improve the QEC performance. Regarding the possible experimental implementation, the proposed tracking QEC will be applicable to repeated quantum nondemolition measurements and adaptive control in a superconducting cavity resonator setup \cite{Hee, Shen, LLi}, which can be regarded as repeated single-qubit level QECs.

The numerical results for the two-QEC cycle showed that the proposed method with analog QEC reduces the required number of the qubits without degrading the QEC performance. The tracking QEC with analog QEC reduces the number of physical qubits required for the $C_{4}/C_{6}$ code by $16\times12^{l-1}-8\times3^{l-1}$ for the $2$-QEC cycle with the concatenation level $l$, and the reduction rate for  is $1/2-1/(4\times4^{l-1})$, where the reduction rates are 25, 43.8, 48.4, 49.6 and 49.9 for the level 1, 2, 3, 4, and 5, respectively.

Furthermore, it has been shown that the analog QEC makes the performances of the single-qubit level QEC almost identical to those of the logical-qubit level QEC under the condition of a practical noise level.
To the best of our knowledge, this approach is the first practical attempt to utilize both the single- and standard logical-qubit level QECs to alleviate the requirement of the number of qubits. Hence, the proposed method has a great advantage in implementing fault-tolerant quantum computation with continuous variables and will open a new way to practical quantum computers.

\section*{Acknowledgements}
This work was funded by ImPACT Program of Council for Science, Technology and Innovation (Cabinet Office, Government of Japan).

\section*{Appendix A:  Logical-qubit level QEC using the $C_{4}/C_{6}$ code}\label{AppA}
\begin{figure}[t]
    \includegraphics[angle=0, width=1.0\columnwidth]{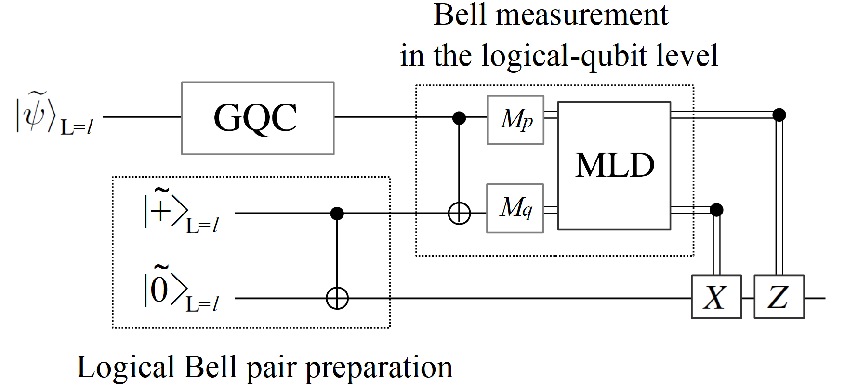}  
    \caption{A quantum circuit for the logical-qubit level QEC based on a quantum teleportation. The logical data qubit $\ket{\widetilde{\psi}}_{{\rm L=}l}$ encoded by the $C_{4}/C_{6}$ code with the concatenation level $l$ is teleported to one of the logical Bell pair, which is prepared from two logical qubits $\ket{\widetilde{+}}_{{\rm L}=l}$ and $\ket{\widetilde{0}}_{{\rm L}=l}$ using the CNOT gate. GQC and  MLD denote the Gaussian quantum channel and a maximum-likelihood decision, respectively.}
 \label{figA1}
\end{figure}

We explain the logical-qubit level QEC using the $C_{4}/C_{6}$ code without and with the analog QEC.
We refer that the conventional logical-qubit level QEC using the $C_{4}/C_{6}$ without the analog QEC is described in Ref. \cite{Goto}. 
In the $C_{4}/C_{6}$ code, the logical-qubit level QEC is based on quantum teleportation as shown in Fig. \ref{figA1}, where the (logical) data qubit  $\ket{\widetilde{\psi}}_{{\rm L={\it l}}}$ with concatenation level $l$ is teleported to one of the fresh (logical) Bell pair $(\ket{\widetilde{0}}_{\rm L={\it l}}\ket{\widetilde{0}}_{\rm L={\it l}}+\ket{\widetilde{1}}_{\rm L={\it l}}\ket{\widetilde{1}}_{\rm L={\it l}})/\sqrt{2}$. 
In quantum teleportation, the Bell measurement in the logical-qubit level is performed on the data qubit and one of the Bell pair. The quantum teleportation process refers to the outcomes $M_{p, {\rm L}=l}$ and $M_{q, {\rm L}=l}$ of the Bell measurement on the encoded qubits with concatenation level $l$. 
After the feedforward operation according to the Bell measurement outcome $M_{p, {\rm L}=l}$ and $M_{q, {\rm L}=l}$, the data is teleported to the other one of Bell pair. If this feedforward is performed correctly, the errors occurred on the data qubit are successfully corrected. 
In the Bell measurement in the physical-qubit level, we obtain the Bell measurement outcomes of bit values $k_{pi}$ and $k_{qi}$ for the $i$-th physical GKP qubit composed of the logical data qubit and the logical Bell state, respectively. In addition to bit values, we also obtain deviation values $\Delta_{p{\rm m}i}$ and $\Delta_{q{\rm m}i}$ for the $i$-th physical GKP qubit, which are used to improve the error tolerance of the Bell measurement by using analog QEC.

As a simple example to describe the logical-qubit level QEC with the analog QEC, we explain the QEC with concatenation level 1, that is, the $C_{4}$ code. The $C_{4}$ code is the $[[\rm{4,2,2}]]$ code and consists of four physical qubits to encode a level-1 qubit pair; thus, it is not the error-correcting code but the error-detecting code in the conventional method. In the measurement of the $C_{4}$ code, we decide the logical bit value $M_{L=1}$ (=0,1) from the bit value $M_{L=1}^1$ and $M_{L=1}^2$ for the first and second qubit of a level-1 qubit pair, respectively. 
The logical bit value of the $C_{4}$ code is $M_{{\rm L}=1}$ when the bit value of the level-1 qubit pair is ($M_{ {\rm L}=1}^1$, 0) or ($M_{{\rm L}=1}^1$ ,1), that is, the bit value of the first qubit $M_{{\rm L}=1}^1$ defines a logical bit value $M_{{\rm L}=1}$ for the level-1. As the parity check of the $Z$ operator for the first and second qubits $ZIZI$ and $IIZZ$ indicates, the bit value of the level-1 qubit pair (0,0) corresponds to the bit value of the physical GKP qubits $(k_{q{\rm1}}, k_{q{\rm 2}}, k_{q{\rm 3}}, k_{q{\rm 4}})$ = (0,0,0,0) or (1,1,1,1)~\cite{Knill}. The bit values of the pairs (0,1), (1,0), and (1,1) correspond to the bit values of the physical GKP qubits (0,1,0,1) or (1,0,1,0), (0,0,1,1) or (1,1,0,0), and (0,1,1,0) or (1,0,0,1), respectively. Therefore, if the measurement outcome is (0,0,1,0) in the $q$ quadrature, we consider two error patterns, assuming the level-1 qubit pair (0,0). The first pattern is a single error on the third physical qubit and the second pattern is the triple errors on the physical qubits except for the third qubit. We then calculate the likelihood for the level-1 qubit pair (0,0) $F_{0,0}$ as
\begin{align}
F_{0 ,0} = f(\sqrt{\pi}-|\Delta_{q{\rm m1}}|)f(\sqrt{\pi}-|\Delta_{q{\rm m2}}|)f(\Delta_{q{\rm m3}})f(\sqrt{\pi}-|\Delta_{q{\rm m4}}|) \nonumber \\ 
                 + f(\Delta_{q{\rm m1}})f(\Delta_{q{\rm m2}})f(\sqrt{\pi}-|\Delta_{q{\rm m3}}|)f(\Delta_{q{\rm m4}}).\hspace{20pt} \tag{A1}
\end{align}
We similarly calculate the $ F_{0,1}, F_{1,0},$ and $F_{1,1}$ likelihood for the bit value of qubit pairs (0,1), (1,0), and (1,1). Finally, we determine the level-1 logical bit value $M_{q, {\rm L}=1}$ in the $q$ quadrature by comparing $F_{0,0}+F_{0,1}$ with $F_{1,0}+F_{1,1}$, which refer to the likelihood functions for the logical bit values zero and one, respectively. If $F_{0,0}+F_{0,1} > F_{1,0}+F_{1,1}$, then we determine $M_{q, {\rm L}=1}$ = 0, and vice versa. Therefore, the analog QEC using likelihood functions can correct the $X$ flip error, that corresponds to the qubit level error in the $q$ quadrature. 
The level-1 logical bit value $M_{p, {\rm L}=1}$ in the $p$ quadrature can be determined by the parity check of the $X$ operator for the first and second qubits $XXII$ and $IXIX$ in a similar manner. In the conventional likelihood method~\cite{Pou,Goto} $F_{0,0}$, $F_{0,1}$, $F_{1,0}$, and $F_{1,1}$ are given by the same joint probability
\begin{equation}
p_{\rm corr}^{3} (1-p_{\rm corr}) + p_{\rm corr}(1-p_{\rm corr})^{3}, \tag{A2}
\end{equation}
where $p_{\rm corr}$ is defined by Eq. (3) in the main text. 
Because $F_{0,0}+F_{0,1} = F_{1,0}+F_{1,1}$, the $C_{4}$ code is not error-correcting code but error-detecting code in the conventional method, whereas it is the error-correcting code in our method. 
The likelihood for the level-$l$ ($l\geqq2$) bit value can be calculated by the likelihood for the level-$(l-1)$ bit value in a similar manner.

\section*{Appendix B: Tracking quantum error correction using the $C_{4}/C_{6}$ code}\label{AppB}
\begin{figure}[t]
    \includegraphics[angle=0, width=1.0\columnwidth]{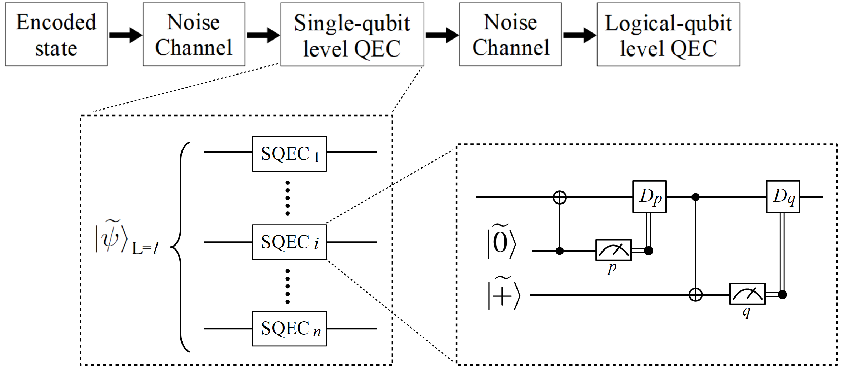}  
    \caption{A quantum circuit for the tracking QEC for two QEC cycles. 
The logical data qubit $\ket{\widetilde{\psi}}_{{\rm L}=l}$ with the concatenation level $l$ is composed of the $n=4 \times 3^{l-1}$ physical qubits. ${\rm SQEC}_{i}$ $(i=1,2, \cdots, n)$ denotes the single-qubit level QEC for the $i$-th qubit, where the single-qubit level QECs in the $p$ and $p$ quadratures are implemented by ancilla physical qubits $\ket{\widetilde{0}}$ and $\ket{\widetilde{+}}$, respectively. }
 \label{figA2}
\end{figure}

We describe the details of the tracking QEC using the $C_{4}/C_{6}$ code with two QEC cycles. 
Fig.~\ref{figA2} shows the tracking QEC in the first cycle, that is, the single-qubit level QECs in the $p$ and $q$ quadratures, where the deviations of physical qubits that constitute a logical data qubit are measured through ancilla qubits and corrected using the displacement operation independently. 
In the first cycle, we obtain the deviation values $\Delta_{p{\rm m}i}^{(1)}$ and $\Delta_{q{\rm m}i}^{(1)}$ for the $i$-th physical qubit in the $q$ and $p$ quadratures, respectively. In the second cycle, we obtain the bit values $k_{p{\rm m}i}$ and $k_{q{\rm m}i}$, and deviation values $\Delta_{p{\rm m}i}^{(2)}$ and $\Delta_{q{\rm m}i}^{(2)}$ in the $p$ and $q$ quadratures, respectively. 
We note that the displacement operation in the single-qubit level QEC is not necessarily for our method, since the displacement operation can be performed in the logical-qubit level QEC all at once. 

As a simple example to describe the tracking QEC, we explain the QEC with concatenation level 1. As described in Appendix A, we decide the logical bit value $M_{q, {\rm L}=1}$ and $M_{p, {\rm L}=1}$ by using the parity check operator obtained from the measurement outcome $k_{q{\rm m}i}$ and $k_{p{\rm m}i}$, respectively. When the measurement outcome ($k_{q{\rm m}1},k_{q{\rm m}2},k_{q{\rm m}3},k_{q{\rm m}4}$ ) is (0,0,1,0) in the $q$ quadrature, we consider two error patterns as described in Appendix A. Considering the two error patterns, where a single error on the third qubit and the triple errors on the physical qubits except for the third qubit, we then calculate the likelihood for the level-1 qubit pair (0,0) $F_{0,0}$ with the analog QEC as
\begin{equation}
F_{0 ,0} = F_{\rm in, 1}^{\rm ana}F_{\rm in, 2}^{\rm ana}F_{\rm corr, 3}^{\rm ana}F_{\rm in, 4}^{\rm ana}
               + F_{\rm corr, 1}^{\rm ana}F_{\rm corr, 2}^{\rm ana}F_{\rm in, 3}^{\rm ana}F_{\rm corr, 4}^{\rm ana}.\tag{B1}
\end{equation}

$F_{\rm corr, {\it i}}^{\rm ana}$ is the likelihood of no error and double errors on the $i$-qubit in the single-qubit level and the logical-qubit level QECs, and described by
\begin{align}
F_{{\rm corr},i}^{\rm ana}=f(|\Delta_{q{\rm m}i}^{(1)}|) f(|\Delta_{q{\rm m}i}^{(2)}|)\hspace{80pt} \nonumber \\
+f(\sqrt{\pi}-|\Delta_{q{\rm m}i}^{(1)}|)f(\sqrt{\pi}-|\Delta_{q{\rm m}i}^{(2)}|). \tag{B2}
\end{align}
 $F_{\rm in, {\it i}}^{\rm ana}$ is the likelihood of a single error on the $i$-qubit in one of the single-qubit level and the logical-qubit level QECs, and described by
\begin{align}
F_{\rm in,{\it i}}^{\rm ana}=f(|\Delta_{q{\rm m}i}^{(1)}|)f(\sqrt{\pi}-|\Delta_{q{\rm m}i}^{(2)}|) \hspace{50pt} \nonumber \\
\hspace{20pt}+f(\sqrt{\pi}-|\Delta_{q{\rm m}i}^{(1)}|)f(|\Delta_{q{\rm m}i}^{(2)}|). \tag{B3}
\end{align}
We calculate the $ F_{0,1}, F_{1,0},$ and $F_{1,1}$ likelihood similarly for the bit value of qubit pairs (0,1), (1,0), and (1,1). In a similar manner of the logical-qubit level QEC, we determine the level-1 logical bit value $M_{q, {\rm L}=1}$ in the $q$ quadrature by comparing $F_{0,0}+F_{0,1}$ with $F_{1,0}+F_{1,1}$. In the tracking QEC without the analog QEC, likelihoods $F_{0,0}$, $F_{0,1}$, $F_{1,0}$, and $F_{1,1}$ are given by the same joint probability
\begin{equation}
F_{\rm corr}^{3} F_{\rm in}+F_{\rm corr} F_{\rm in}^3, \tag{B4}
\end{equation}
where $F_{\rm corr}$ and $F_{\rm in}$ are defined by Eqs. (21) and (23) in the main text, respectively. 
Therefore, the tracking QEC without the analog QEC is not error-correcting code but error-detecting code, whereas that with the analog QEC is the error-correcting code. The likelihood for the level-$l$ ($l\geqq2$) bit value can be calculated by the likelihood for the level-$(l-1)$ bit value in a similar manner.



\end{document}